\documentclass[12pt]{article}

\usepackage{graphicx}
\begin{document}

\begin{center}
{\bf Notes on Born$-$Infeld-type electrodynamics } \\
\vspace{5mm} S. I. Kruglov
\footnote{E-mail: serguei.krouglov@utoronto.ca}
\underline{}
\vspace{3mm}

\textit{Department of Chemical and Physical Sciences, University of Toronto,\\
3359 Mississauga Road North, Mississauga, Ontario L5L 1C6, Canada} \\
\vspace{5mm}
\end{center}
\begin{abstract}
We propose a new model of nonlinear electrodynamics with three parameters. Born$-$Infeld electrodynamics and exponential electrodynamics are particular cases of this model. The phenomenon of vacuum birefringence in the external magnetic field is studied. We show that there is not singularity of the electric field at the origin of point-like charged particles. The corrections to Coulomb's law at $r\rightarrow\infty$ are obtained. We calculate the total electrostatic energy of charges, for different parameters of the model, which is finite.
\end{abstract}

\section{Introduction}

For strong electromagnetic fields classical electrodynamics is modified due to self-interaction of photons \cite{Jackson}.
Quantum electrodynamics (QED) with loop corrections becomes nonlinear electrodynamics (NLED) \cite{Heisenberg} and the effect of vacuum birefringence occurs in the presence of the external magnetic field. This phenomenon is now of experimental verification \cite{Rizzo}, \cite{Valle}, \cite{Battesti}.
In Born$-$Infeld (BI) electrodynamics \cite{Born} there is no the effect of birefringence in the background magnetic field but in some NLED models the phenomenon of vacuum birefringence in the external magnetic field takes place \cite{Kruglov9}-\cite{Kruglov0}.
Some models of NLED coupled with general relativity were investigated in \cite{Kruglov0}-\cite{Balart} and in many other papers.
The correspondence principle requires that
in the limit of weak fields NLED has to be converted into Maxwell's electrodynamics. In classical electrodynamics
problems of singularity of an electric field in the center of particles and the infinite electromagnetic energy of charges exist.
The attractive feature of some NLED models \cite{Born}, \cite{Kruglov2}, \cite{Kruglov}, \cite{Shabad}, \cite{Kruglov5} is that there is an upper bound on the electric field at the origin of charges and the total electrostatic energy is finite. In the models \cite{Shabad1}-\cite{Kruglov6} 
the electric field at the centre of charges is infinite but self-energy of particles is finite.
In this paper we propose a new model of NLED with tree parameters, where the singularity is absent and the total electrostatic energy is finite. The model under consideration is parameterized by three
parameters that contains the known BI model, the exponential model, and its special cases. The Lagrangian density of the model under investigation can also match the Heisenberg-Euler (HE) Lagrangian density within
the lowest terms of the expansion in powers of fields. The comparison of the parameters with QED is done up to the
two-loop level of the HE Lagrangian. One can consider this as a motivation for Lagrangian density (1).

The paper is organised as follows. In section 2 we formulate the new model of NLED. The phenomenon of vacuum birefringence in the background magnetic field is discussed in subsection 2.1. We obtain the bound on the parameters of the model from
the BMV and PVLAS experiments. In section 3 we briefly discuss the field equations of NLED proposed. The Coulomb low corrections are calculated in subsection 3.2. We found the maximum of electric
field strength at the centre of point-like charged particles. The self-energy of charges is calculated in section 4 for different parameters of the model. Section 5 is devoted to a conclusion. In Appendix we show that the phenomenon of vacuum birefringence in background field is absent in model only approximately if $\beta\neq\gamma$, $\sigma\neq 0.5$. Otherwise, when  $\beta =\gamma$, $\sigma = 0.5$, we come to BI electrodynamics, and the effect of vacuum birefringence in background field disappears exactly in all orders $\beta{\cal F}$ and $\gamma{\cal G}$.

We use units with $c=\hbar=1$, $\varepsilon_0=\mu_0=1$, and the metric signature is given by $\eta=\mbox{diag}(-1,1,1,1)$.

\section{A new model of nonlinear electrodynamics}

Let us consider Born$-$Infeld-type electrodynamics with the Lagrangian density
\begin{equation}
{\cal L} = \frac{1}{\beta}\left[1-\left(1+\frac{\beta{\cal F}}{\sigma}-\frac{\beta\gamma {\cal G}^2}{2\sigma}\right)^\sigma\right],
 \label{1}
\end{equation}
where the parameters $\beta$ and $\gamma$ have the dimensions of (length)$^4$, and $\sigma$ is the dimensionless parameter, ${\cal F}=(1/4)F_{\mu\nu}F^{\mu\nu}=(\textbf{B}^2-\textbf{E}^2)/2$, ${\cal G}=(1/4)F_{\mu\nu}\tilde{F}^{\mu\nu}=\textbf{E}\cdot \textbf{B}$,
$F_{\mu\nu}=\partial_\mu A_\nu-\partial_\nu A_\mu$ is the field strength tensor, and $\tilde{F}^{\mu\nu}=(1/2)\epsilon^{\mu\nu\alpha\beta}F_{\alpha\beta}$
is the dual tensor. When $\sigma=1/2$ the model (1) is converted into the model introduced in \cite{Krug}.
At $\beta=\gamma$ and $\sigma=1/2$ we arrive at BI electrodynamics.
The model described by the Lagrangian density (1) obeys the correspondence principle, i.e. at the weak field limit, $\beta {\cal F}\ll 1$, $\gamma{\cal G}\ll 1$, NLED (1) is converted into Maxwell'l electrodynamics, ${\cal L}_M=-{\cal F}$.
As a result, the nonlinearity of field equations is absent in the weak field limit.
It should be noted that in the Born$-$Infeld-like model considered in \cite{Helael} the correspondence principle does not hold for any parameter $p$ introduced. At $\sigma\rightarrow \infty$ the model (1) is transformed into exponential electrodynamics
\begin{equation}
{\cal L}_{exp}=\lim_{\sigma\rightarrow\infty}{\cal L} = \frac{1}{\beta}\left[1-\exp\left(\beta{\cal F}-\beta\gamma {\cal G}^2/2\right)\right].
 \label{2}
\end{equation}
Other variants of exponential electrodynamics were proposed in \cite{Hendi}, \cite{Kruglov0}. So, our model (1) allows us to consider simultaneously BI and exponential electrodynamics, as well as a new version of NLED.

The interest in BI model, which is a particular case of the model (1), is in the fact that the low energy D-brain dynamics is governed by BI-type action \cite{Fradkin}, \cite{Tseytlin}. We treat Eq. (1) proposed as an effective Lagrangian density which takes into account quantum corrections and is valid for strong electromagnetic fields when the perturbation method does not work. This situation can be realised, for example, in early time of universe creation (Big Bang) or inside of charged black holes.

\subsection{Vacuum birefringence in the background magnetic field}

Because in QED the effect of vacuum birefringence in the external magnetic field occurs \cite{Heisenberg}, due to quantum corrections, it is of interest to study this effect in the model (1). The BMV experiment \cite{Rizzo} and the PVLAS experiment \cite{Valle} provided bounds on the magnitude of the phenomenon of vacuum birefringence in the background magnetic field. In classical electrodynamics and in BI \cite{Born} electrodynamics vacuum birefringence in the background magnetic field is absent but
in generalized BI electrodynamics with two parameters \cite{Krug} the phenomenon of vacuum birefringence in the external magnetic field takes place. In the weak field limit $\beta {\cal F}\ll 1$, $\gamma {\cal G}\ll 1$, the Lagrangian density (1), making use of the Taylor series, becomes
\[
{\cal L} = -{\cal F}+\frac{\beta (1-\sigma)}{2\sigma}{\cal F}^2+\frac{\gamma}{2}{\cal G}^2
\]
\begin{equation}
-\frac{\beta^2 (\sigma-1)(\sigma-2)}{6\sigma^2}{\cal F}^3+\frac{\beta \gamma(\sigma-1)}{2\sigma}{\cal F}{\cal G}^2
+{\cal O}\left((\beta {\cal F})^4,(\gamma{\cal G})^4\right).
 \label{3}
\end{equation}
Up to ${\cal O}\left((\beta{\cal F})^2,(\gamma{\cal G})^2\right)$ the Lagrangian density (3) becomes the Heisenberg-Euler type \cite{Kruglov9}
\begin{equation}
{\cal L}_0 =- \frac{1}{2}\left(\textbf{B}^2-\textbf{E}^2\right)+a\left(\textbf{B}^2-\textbf{E}^2\right)^2
+b\left(\textbf{E}\cdot\textbf{B}\right)^2,
\label{4}
\end{equation}
where $a=\beta (1-\sigma)/8\sigma$, $b=\gamma/2$. The phenomenon of vacuum birefringence (in the background magnetic field) in the model with the Lagrangian density (4) was investigated in \cite{Kruglov9} (see also \cite{Dittrich}). It was shown that the indices of refraction $n_\perp$, $n_\|$ for two polarizations, perpendicular and parallel to the external magnetic induction field $\bar{B}$, are
\begin{equation}
n_\perp=1+4a\bar{B}^2=1+\frac{\beta(1-\sigma)}{2\sigma}\bar{B}^2,~~~~n_\|=1+b\bar{B}^2=1+\frac{\gamma}{2}\bar{B}^2.
\label{5}
\end{equation}
The phase velocities become $v_\perp=1/n_\perp$, $v_\|=1/n_\|$ and the phenomenon of vacuum birefringence in the external magnetic field holds when $n_\perp\neq n_\|$.
According to the Cotton-Mouton (CM) effect \cite{Battesti} the $\triangle n_{CM}$ reads
\begin{equation}
\triangle n_{CM}=n_\|-n_\perp=k_{CM}\bar{B}^2.
\label{6}
\end{equation}
With the aid of Eqs. (5) and (6) one finds $k_{CM}=\gamma/2-\beta (1-\sigma)/2\sigma$.
From the BMV and PVLAS experiments
\[
k_{CM}=(5.1\pm 6.2)\times 10^{-21} \mbox {T}^{-2}~~~~~~~~~~(\mbox {BMV}),
\]
\begin{equation}
k_{CM}=(4\pm 20)\times 10^{-23} \mbox {T}^{-2}~~~~~~~~~~~~(\mbox {PVLAS}),
\label{7}
\end{equation}
we obtain the bound on our model parameters $\gamma/2-\beta(1-\sigma)/2\sigma\leq(4\pm 20)\times 10^{-23}\mbox {T}^{-2}$.
The effect of vacuum birefringence in the background magnetic field is absent if $k_{CM}=0$ ($\gamma=\beta (1-\sigma)/\sigma$). 
It should be noted that this relation, that not present birefringence, was based on the analyses of approximate Lagrangian density (3) up to ${\cal O}((\beta{\cal F})^2)$, ${\cal O}((\gamma{\cal G})^2)$, and, therefore, it holds approximately. It is proven in Appendix, by means of the method given in \cite{Melo}, that the effect of vacuum birefringence disappears in our model only approximately when $\gamma=\beta(1-\sigma)/\sigma$, $\sigma\neq0.5$. Our approximation is justified because the experimental difference in indices of refraction (6) is expressed through quadratic external magnetic field $\bar{B}^2$ or linear in $\bar{{\cal F}}$. Only in BI electrodynamics when $\gamma=\beta$, $\sigma=0.5$ the vacuum birefringence effect in the external magnetic field is absent exactly in all order in $\beta{\cal F}$ without approximations. In exponential electrodynamics, with the Lagrange density (2), when $\sigma\rightarrow\infty$ the phenomenon of vacuum birefringence (in the external magnetic field) disappears approximately if $\beta=-\gamma$.
The bound on CM coefficient in QED is given by $k_{CM}\leq 4.0\times 10^{-24}\mbox {T}^{-2}$ \cite{Rizzo}.

For weak fields we come to QED and quantum corrections lead to the Heisenberg-Euler action. Two-loop
corrected Heisenberg-Euler Lagrangian density reads as Eq. (4) with the parameters \cite{Ritus}, \cite{Ritus1}, \cite{Dittrich1}
\begin{equation}\label{8}
  a=\frac{4\alpha^2}{45m^4}\left(1+\frac{40\alpha}{9\pi}\right),~~~~
b=\frac{14\alpha^2}{45m^4}\left(1+\frac{1315\alpha}{252\pi}\right),
\end{equation}
where $m$ is the electron mass, and $\alpha=e^2/(4\pi)\approx 1/137$. Then we identify $\beta (1-\sigma)/8\sigma=a$ and $\gamma/2=b$. Thus, the parameter $\gamma$ is fixed and another bound on the parameters $\beta$ and $\sigma$ follows from the birefringence experiment. As a result, Eq. (1) gives a convenient parametrization that allows us to study different models.

\section{Field equations}

Making use of Euler-Lagrange equations we obtain equations of motion
\begin{equation}
\partial_\mu\left({\cal L}_{\cal F}F^{\mu\nu} +{\cal L}_{\cal G}\tilde{F}^{\mu\nu} \right)=0,
\label{9}
\end{equation}
where ${\cal L}_{\cal F}=\partial {\cal L}/\partial{\cal F}$, ${\cal L}_{\cal G}=\partial {\cal L}/\partial{\cal G}$, and $\tilde{F}^{\mu\nu}=(1/2)\epsilon^{\mu\nu\alpha\beta}F_{\alpha\beta}$ is the dual tensor.
By virtue of Eqs. (1) and (9) one finds field equations
\begin{equation}
 \partial_\mu\left[\Pi^{\sigma-1}\left(F^{\mu\nu}-\gamma{\cal G}\tilde{F}^{\mu\nu}\right) \right]=0,~~~~~
\Pi=1+\frac{\beta{\cal F}}{\sigma}-\frac{\beta\gamma {\cal G}^2}{2\sigma}.
\label{10}
\end{equation}
The electric displacement field is given by $\textbf{D}=\partial{\cal L}/\partial \textbf{E}$,
\begin{equation}
\textbf{D}=\Pi^{\sigma-1}\left( \textbf{E}+\gamma {\cal G}\textbf{B}\right).
\label{11}
\end{equation}
The magnetic field obtained from the relation $\textbf{H}=-\partial{\cal L}/\partial \textbf{B}$ is
\begin{equation}
\textbf{H}=\Pi^{\sigma-1}\left( \textbf{B}-\gamma{\cal G}\textbf{E}\right).
\label{12}
\end{equation}
We can decompose Eqs. (11) and (12) as \cite{Hehl}
\begin{equation}
\textbf{D}=\varepsilon \textbf{E}+\nu \textbf{B},~~~~\textbf{H}=\mu^{-1}\textbf{B}-\nu \textbf{E}.
\label{13}
\end{equation}
From Eqs. (11), (12) and (13) one finds
\begin{equation}
\varepsilon=\Pi^{\sigma-1},~~~~
\mu^{-1}=\varepsilon,~~~~\nu=\gamma {\cal G}\Pi^{\sigma-1}.
\label{14}
\end{equation}
With the help of Eqs. (11) and (12), field equations (9) may be written in the form of nonlinear Maxwell's equations
\begin{equation}
\nabla\cdot \textbf{D}= 0,~~~~ \frac{\partial\textbf{D}}{\partial
t}-\nabla\times\textbf{H}=0.
\label{15}
\end{equation}
As $\varepsilon$, $\mu^{-1}$, and $\nu$ depend on electromagnetic fields, equations (15) are nonlinear Maxwell's equations.
The Bianchi identity gives the second pair of nonlinear Maxwell's equations
\begin{equation}
\nabla\cdot \textbf{B}= 0,~~~~ \frac{\partial\textbf{B}}{\partial
t}+\nabla\times\textbf{E}=0.
\label{16}
\end{equation}
Making use of Eqs. (13) and (14) we obtain the equality
\begin{equation}
\textbf{D}\cdot\textbf{H}=\textbf{E}\cdot\textbf{B}\left(1+2\gamma{\cal F}-\gamma^2{\cal G}^2\right)\Pi^{2(\sigma-1)}.
\label{17}
\end{equation}
It follows from Eq. (17) that $\textbf{D}\cdot\textbf{H}\neq\textbf{E}\cdot\textbf{B}$ if $\sigma\neq 0.5$ or $\beta\neq\gamma$, and therefore (see \cite{Gibbons}), the dual symmetry is broken in our model for this case as well as in generalized BI electrodynamics \cite{Krug} and in QED with quantum corrections. But when $\sigma =0.5$ and $\beta=\gamma$, we come to BI electrodynamics and the dual symmetry holds as well as in classical electrodynamics.

\subsection{The electric field of point-like charged particles}

Adding the source in Eq. (15) for the point-like charged particle with the electric charge $Q$, in Gaussian units, we obtain the equation
\begin{equation}
\nabla\cdot \textbf{D}=4\pi Q\delta(\textbf{r}).
\label{18}
\end{equation}
 The solution to Eq. (18) is
\begin{equation}
 \textbf{D}=\frac{Q}{r^3}\textbf{r}.
\label{18}
\end{equation}
By virtue of Eq. (11) it becomes
\begin{equation}
\frac{E}{[1-\beta E^2/(2\sigma)]^{1-\sigma}}=\frac{Q}{r^2}.
\label{20}
\end{equation}
We make a conclusion that if $\sigma < 1$
at $r\rightarrow 0$ the solution to Eq. (20) is given by
\begin{equation}
E(0)=\sqrt{\frac{2\sigma}{\beta}}.
\label{21}
\end{equation}
At $\sigma=1/2$ we come to known result that in BI electrodynamics the maximum field in the origin of charged particle is $E(0)=\sqrt{1/\beta}$. For the case of $\sigma>1$, as in exponential electrodynamics ($\sigma\rightarrow\infty$), there is a singularity in the center of particles.
At $\sigma<1$, there is no singularity of the electric field strength at the origin of point-like charged particles and
Eq. (21) gives the maximum of the electric field in the center of charged particles.
Let us introduce unitless variables
\begin{equation}
x=\frac{(2\sigma)^{1/4}}{\sqrt{Q}\beta^{1/4}}r,~~~~y=\sqrt{\frac{\beta}{2\sigma}}E.
\label{22}
\end{equation}
Using Eqs. (22), equation (20) takes the form
\begin{equation}
\frac{y}{(1-y^2)^{1-\sigma}}=\frac{1}{x^2}.
\label{23}
\end{equation}
The plots of the function $y(x)$ for $\sigma=0.1$, $\sigma=0.5$ and $\sigma=0.8$ are represented by Fig. 1.
\begin{figure}[h]
\includegraphics[height=3.0in,width=3.0in]{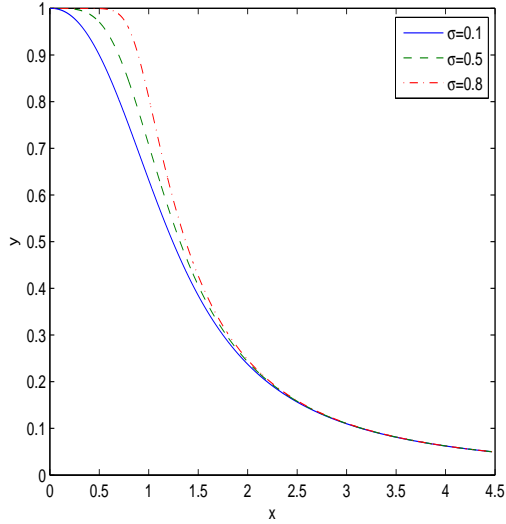}
\caption{The plot of the function $y$ vs. $x$.}
\label{fig.1}
\end{figure}

The Taylor series of the function $y(x)$ at $x\rightarrow \infty$ ($r\rightarrow \infty$) reads
\begin{equation}
y=\frac{1}{x^2}-\frac{1-\sigma}{x^6}+{\cal O}(x^{-8}).
\label{24}
\end{equation}
From Eqs. (22) and (24) one can obtain the asymptotic value of the electric field  at $r\rightarrow\infty$
\begin{equation}
E(r)=\frac{Q}{r^2}-\frac{(1-\sigma)\beta Q^3}{2\sigma r^6}+{\cal O}(r^{-8}).
\label{25}
\end{equation}
Eq. (25) shows the correction to Coulomb's law at $r\rightarrow\infty$ for any value of the parameter $\sigma$.
For BI electrodynamics, when $\sigma=1/2$, we find the asymptotic of the electric field
\begin{equation}
E(r)=\frac{Q}{r^2}-\frac{\beta Q^3}{2 r^6}+{\cal O}(r^{-8})~~~~~~(\sigma=1/2).
\label{26}
\end{equation}
For exponential electrodynamics ($\sigma\rightarrow \infty$) one obtains from Eq. (25) the correction to Coulomb's law similar to
BI electrodynamics but with the opposite sign, $E(r)=Q/r^2+\beta Q^3/(2 r^6)+{\cal O}(r^{-8})$.
At $\beta=0$ we come to Maxwell's electrodynamics and the Coulomb law $E(r)=Q/r^2$ is recovered.
The electric field singularities are absent at the origin of charged particles at $\sigma<1$.

\section{Energy-momentum tensor and the energy of charged particles}

The general expression for the symmetrical energy-momentum tensor is given by
\begin{equation}
T_{\mu\nu}=\left({\cal L}_{\cal F}F_\mu^{~\alpha}+{\cal L}_{\cal G}\tilde{F}_\mu^{~\alpha}\right)F_{\nu\alpha}-g_{\mu\nu}{\cal L}.
\label{27}
\end{equation}
From Eqs. (1) and (27) we find the symmetrical energy-momentum tensor
\begin{equation}
T_{\mu\nu}=-\Pi^{\sigma-1}\left(F_\mu^{~\alpha}F_{\nu\alpha}-
\gamma{\cal G}\tilde{F}_\mu^{~\alpha}F_{\nu\alpha}\right)-g_{\mu\nu}{\cal L}.
\label{28}
\end{equation}
For the case $\textbf{B}=0$ the electric energy density becomes
\begin{equation}
\rho_E=T^{~0}_0=\frac{E^2}{(1-\beta E^2/(2\sigma))^{1-\sigma}}+\frac{1}{\beta}\left[\left(1-\frac{\beta E^2}{2\sigma}\right)^{\sigma}-1\right].
\label{29}
\end{equation}
The total electrostatic energy of point-like charges is given by ${\cal E}=\int_0^\infty \rho_E r^2dr$. Making use of Eqs. (22), (23) and (29), and changing variables, we obtain
\[
{\cal E}=\frac{Q^{3/2}}{2^{7/4}\sigma^{3/4}\beta^{1/4}}\int_0^1 \left[1-\frac{2\sigma y^2}{(1-y^2)^{1-\sigma}}-(1-y^2)^\sigma\right]
\]
\begin{equation}
\times\frac{(1-y^2)^{(1-3\sigma)/2}\left[(2\sigma-1)y^2-1\right]dy}{y^{5/2}}.
\label{30}
\end{equation}
The normalized electrostatic energy ${\cal E}\beta^{1/4}/Q^{3/2}$ of point-like charges for different parameters $\sigma$, found from Eq. (30), is given in Table 1.
\begin{table}[ht]
\caption{The normalized electrostatic energy}
\centering
\begin{tabular}{c c c c c c c c c }\\[1ex]
\hline \hline
$\sigma$ & 0.1 & 0.2 & 0.3 & 0.4 & 0.5 & 0.6 & 0.7 & 0.8  \\[0.5ex]
\hline
${\cal E}\beta^{1/4}/Q^{3/2}$ & 0.788 & 0.947 & 1.061 & 1.154 & 1.250 & 1.311 & 1.382 & 1.451 \\[0.5ex]
\hline
\end{tabular}
\end{table}
Thus, the total electrostatic energy of point-like particles is finite for $0<\sigma<1$.

 \section{Conclusion}

We have proposed a new model of NLED with three parameters $\beta$, $\gamma$ and $\sigma$ and BI electrodynamics (at $\beta= \gamma$, $\sigma=1/2$) and exponential electrodynamics (at $\sigma\rightarrow\infty$) are particular cases of this model. The correspondence principle takes place so that for weak fields the model is converted to Maxwell's electrodynamics. The phenomenon of vacuum birefringence  in the background magnetic field occurs if $\gamma\neq\beta(1-\sigma)/\sigma$ otherwise the effect of birefringence disappears approximately if we imply that $\beta{\cal F}\ll1$, $\gamma{\cal G}\ll1$. Otherwise, when  $\beta=\gamma$, $\sigma=0.5$, we come to BI electrodynamics, and the effect of vacuum  birefringence disappears exactly in all orders of $\beta{\cal F}$ and  $\gamma{\cal G}$.
We demonstrated that the dual symmetry is violated in our model if $\sigma\neq 0.5$ or $\beta\neq\gamma$.
It is shown that there is no singularity of the electric field in the center of point-like charged particles and the maximum electric field strength is $E(0)=\sqrt{2\sigma}/\sqrt{\beta}$. The corrections to Coulomb's law at $r\rightarrow\infty$ are obtained which are in the order of ${\cal O}(r^{-6})$. We calculated the total electrostatic energy of charged particles for different values of the parameter $\sigma$, that is finite.

 \section{Appendix}
 
The analysis of no-birefringence was given in \cite{Melo} based on works of \cite{Boillat} and \cite{Birula} (see also \cite{Novello1}). It was shown in \cite{Melo} that phenomenon of vacuum  birefringence in the external field is absent if relations
\begin{equation}
{\cal G}({\cal L}^2_{{\cal F}{\cal G}}-{\cal L}_{{\cal F}{\cal F}}{\cal L}_{{\cal G}{\cal G}})+{\cal L}_{\cal F}{\cal L}_{{\cal F}{\cal G}}=0,
\label{31}
\end{equation}
\begin{equation}
2{\cal F}({\cal L}^2_{{\cal F}{\cal G}}-{\cal L}_{{\cal F}{\cal F}}{\cal L}_{{\cal G}{\cal G}})+
{\cal L}_{\cal F}({\cal L}_{{\cal F}{\cal F}}-{\cal L}_{{\cal G}{\cal G}})=0
\label{32}
\end{equation}
hold. Here $F_{\mu\nu}$ is a background constant electromagnetic field strength tensor and  ${\cal F}=(1/4)F_{\mu\nu}F^{\mu\nu}$, ${\cal G}=(1/4)F_{\mu\nu}\tilde{F}^{\mu\nu}$. From Eq. (1) we obtain
\[
{\cal L}_{\cal F}=-\Pi^{\sigma-1},~~~{\cal L}_{{\cal F}{\cal F}}=\frac{(1-\sigma)\beta}{\sigma}\Pi^{\sigma-2},~~~
{\cal L}_{{\cal F}{\cal G}}=\frac{(\sigma-1)\beta\gamma}{\sigma}{\cal G}\Pi^{\sigma-2},
\]
\begin{equation}
{\cal L}_{\cal G}=\gamma{\cal G}\Pi^{\sigma-1},~~~{\cal L}_{{\cal G}{\cal G}}=\left(\gamma\Pi-\frac{(\sigma-1)\beta\gamma^2}{\sigma}{\cal G}^2\right)\Pi^{\sigma-2},
\label{33}
\end{equation}
where $\Pi$ is given by Eq. (10). Making use of Eq. (33), one can verify that Eq. (31) is valid for any parameters $\beta$, $\gamma$ and $\sigma$. Inserting expressions (33) into Eq. (32) we obtain the equation as follows:
\begin{equation}
\frac{(\sigma-1)\beta}{\sigma}+\gamma+\frac{(2\sigma-1)\beta\gamma}{\sigma}{\cal F}-\frac{(2\sigma-1)\beta\gamma^2}{2\sigma}{\cal G}^2=0.
\label{34}
\end{equation}
Equation (34) is satisfied for $\beta=\gamma$, $\sigma=0.5$, i.e. for BI electrodynamics. if we neglect small values $\beta{\cal F}\ll 1$, $\gamma{\cal G}\ll 1$, then Eq. (34) is valid if the link between parameters Holds: 
\begin{equation}
\gamma=\frac{(1-\sigma)\beta}{\sigma}.
\label{35}
\end{equation}
Thus, the phenomenon of vacuum birefringence in background field is absent in our model only approximately  (if $\beta\neq\gamma$, $\sigma\neq 0.5$) when the condition (35) occurs. In BI electrodynamics, $\beta=\gamma$, $\sigma=0.5$, Eqs. (34) and (35) hold exactly for any values of external field invariants ${\cal F}$ and ${\cal G}$. This result is consistent with works of \cite{Boillat}, \cite{Plebanski} and \cite{Deser}.

\end{document}